\RequirePackage{silence}
\documentclass[aps,pra,twocolumn,showpacs,amsmath,amssymb,preprintnumbers,superscriptaddress,10pt,longbibliography]{revtex4-2}

\usepackage{graphicx}
\graphicspath{{figures/}}
\usepackage{SIunits}
\usepackage{multirow}
\usepackage{makecell}
\usepackage{hyphenat}
\usepackage[bookmarks=false]{hyperref}
\usepackage{color}
\usepackage[capitalise]{cleveref}

\crefname{section}{Sec.}{Secs.}% APS style uses abbreviations
\Crefname{section}{Section}{Sections}

\definecolor{darkorchid}{RGB}{153,50,204}

\definecolor{cardinal}{RGB}{196,30,58}

\definecolor{blue}{RGB}{0,0,255}

\definecolor{cobalt}{RGB}{0,71,171}

\definecolor{teal}{RGB}{0,128,128}

\definecolor{pumpkin}{RGB}{255, 117, 24}

\definecolor{pink}{RGB}{255,0,255}

\begin{document}

\title{Robustness of fiber-optic attenuators to 1061-nm sub-nanosecond pulsed laser radiation in quantum key distribution systems}

\author{Daria~Ruzhitskaya}
\affiliation{Russian Quantum Center, Skolkovo, Moscow 121205, Russia}
\affiliation{NTI Center for Quantum Communications, National University of Science and Technology MISiS, Moscow 119049, Russia}

\author{Irina~Zhluktova}
\affiliation{Prokhorov General Physics Institute of Russian Academy of Sciences, Moscow 119991, Russia}

\author{Anastasiya~Ponosova}
\affiliation{Russian Quantum Center, Skolkovo, Moscow 121205, Russia}
\affiliation{NTI Center for Quantum Communications, National University of Science and Technology MISiS, Moscow 119049, Russia}
\affiliation{Prokhorov General Physics Institute of Russian Academy of Sciences, Moscow 119991, Russia}

\author{Fedor~Ushakov}
\affiliation{Prokhorov General Physics Institute of Russian Academy of Sciences, Moscow 119991, Russia}
\affiliation{National Research Nuclear University MEPhI, Moscow 115409, Russia}

\author{Andrey~Zverev}
\affiliation{Prokhorov General Physics Institute of Russian Academy of Sciences, Moscow 119991, Russia}
\affiliation{Ulyanovsk State University, Ulyanovsk, 432970, Russia}

\author{Galina~Tertyshnikova}
\affiliation{Ulyanovsk State University, Ulyanovsk, 432970, Russia}

\author{Tianyi~Xing}
\affiliation{College of Computer Science and Technology, National University of Defense Technology, Changsha 410073, China}

\author{Kirill~Min'kov}
\affiliation{Russian Quantum Center, Skolkovo, Moscow 121205, Russia}

\author{Daniil~Trefilov}
\affiliation{Vigo Quantum Communication Center, University of Vigo, Vigo E-36310, Spain}
\affiliation{School of Telecommunication Engineering, Department of Signal Theory and Communications, University of Vigo, Vigo E-36310, Spain}
\affiliation{atlanTTic Research Center, University of Vigo, Vigo E-36310, Spain}

\author{Anqi~Huang}
\affiliation{College of Electronic Science and Technology, National University of Defense Technology, Changsha 410073, China}

\author{Vladimir~Kamynin}
\affiliation{Prokhorov General Physics Institute of Russian Academy of Sciences, Moscow 119991, Russia}

\author{Vladimir~Tsvetkov}
\affiliation{Prokhorov General Physics Institute of Russian Academy of Sciences, Moscow 119991, Russia}

\author{Vadim~Makarov}
\affiliation{Russian Quantum Center, Skolkovo, Moscow 121205, Russia}
\affiliation{Vigo Quantum Communication Center, University of Vigo, Vigo E-36310, Spain}
\affiliation{NTI Center for Quantum Communications, National University of Science and Technology MISiS, Moscow 119049, Russia}

\date{\today}

\begin{abstract}
The security of quantum key distribution (QKD) systems relies on the physical integrity of their components. While laser-damage attacks (LDAs) using high-power continuous-wave (cw) lasers have been well studied, the threat posed by pulsed lasers at alternative wavelengths remains underestimated. Here, we experimentally investigated the stability of four types of fiber-optic attenuators under exposure to sub-picosecond pulses at $1061~\nano\meter$ with average power reaching $1~\watt$. Mechanical variable attenuators with blocking elements and fixed air-gap attenuators show resistance to this attack. MEMS-based variable attenuators exhibit increased attenuation or irreversible damage that causes a permanent reduction in attenuation of approximately $3.8~\deci\bel$. For fixed attenuators with an absorption element, we demonstrate that initial pulsed irradiation significantly lowers the optical damage threshold of the components compared to direct cw attacks. The attenuation reduction achieved is up to $7~\deci\bel$ at a $1~\watt$ cw laser at $1550~\nano\meter$. These results highlight the possibility of establishing a hidden side-channel for eavesdropping attacks and underscore the insufficiency of existing countermeasures against sophisticated LDA scenarios.
\end{abstract}

\maketitle

\section{Introduction}
\label{sec:intro}

Quantum key distribution (QKD) systems play an important role in modern secure communication by using fundamental principles of quantum mechanics to generate cryptographic keys resistant to computational attacks~\cite{bennett1984, lo2014}. Despite their theoretical reliability~\cite{shor2000, koashi2009}, practical QKD implementations remain vulnerable to side-channel attacks arising from hardware imperfections~\cite{sajeed2021, sun2022, makarov2024, marquardt2023}. These imperfections can be amplified or even intentionally created by an adversary via light injection. A critical threat in this context is the laser-damage attack (LDA), where an eavesdropper Eve introduces high-power laser radiation into the quantum channel to change the operating parameters of the system's optical components~\cite{bugge2014, makarov2016, huang2020, ponosova2022, han2023}.

Most existing studies in this field focus on LDAs using continuous-wave (cw) lasers operating at the standard telecommunication wavelength of $1550~\nano\meter$~\cite{huang2020, ponosova2022, bugge2014, makarov2016, bugai2022}. In particular, such radiation can compromise certain types of fiber-optic attenuators~\cite{huang2020, bugai2022} that reduce QKD signal pulses to a single-photon level and guarantee secure key exchange. Experiments revealed that while some attenuator types (e.g., variable optical attenuators with a blocking element) resist $1550$-$\nano\meter$ cw LDAs, others fail under high-power laser exposure, showing a drop in attenuation and an uncontrolled rise in mean photon number~\cite{huang2020, bugai2022}. Understanding these effects enables QKD developers to mitigate these risks by positioning the proper attenuator types directly at Alice’s exit~\cite{pljonkin2017,kurochkin2024} or after a protective isolator~\cite{ponosova2022}.
%In particular, fiber-optic attenuators, which are essential components of the source in QKD systems, can be damaged by such radiation~\cite{huang2020, bugai2022}. This damage leads to an uncontrolled increase in the mean photon number of transmitted states, creating a loophole for secret keys interception. 
%These components are typically positioned in the QKD source immediately after the protective isolator, making them the next potential target for attacks that could bypass the first line of defense~\cite{huang2020, tan2025}. 
%These components reduce the intensity of QKD signal pulses to a single-photon level, guaranteeing the security of key exchange. Attenuators might be positioned in the QKD source directly at Alice's exit~\cite{pljonkin2017,kurochkin2024}, as several attenuator types have demonstrated high resilience against the $1550~\nano\meter$  cw LDA~\cite{huang2020}. Alternatively, they can be placed immediately after the protective isolator, making them the next potential target for attacks in case of bypassing the first line of defence~\cite{huang2020, tan2025}. 
However, these studies~\cite{huang2020, bugai2022} do not cover the full spectrum of possible LDA scenarios. A comprehensive security analysis, guided by Kerckhoffs's principle, must assume that the adversary has full knowledge of the system's physical implementation and will employ the most effective attack strategy available~\cite{kerckhoffs1883}. This implies that Eve is not limited in her choice of laser parameters, including operating wavelength and working regime, and may select those that are most damaging or hardest to detect.
	
To fully understand LDA possibilities, it is crucial to consider that attack effectiveness can be significantly increased by using pulsed lasers (PLs) and alternative wavelengths. Unlike cw radiation, short high-energy pulses can induce nonlinear processes and electronic excitation in optical component materials~\cite{wood2003, kang2025, Manenkov2014, stuart1995, peng2024}. This leads to more efficient or qualitatively different damage, even at low average levels of optical power. In addition, the efficiency of LDA can be further improved by using lasers with wavelengths shorter than $1550~\nano\meter$, which is the wavelength commonly used in QKD implementations~\cite{makarov2024, sajeed2021}. The laser-induced damage threshold for typical optical materials decreases at shorter wavelengths, primarily due to the increased probability of multiphoton absorption processes~\cite{Karr2003, Chambonneau2018, wood2003}. This means that a lower average optical power is necessary to modify the properties of a component compared to an attack at $1550~\nano\meter$~\cite{lucamarini2015, Karr2003, Chambonneau2018}.
	
Our previous research confirmed this hypothesis: we demonstrated that fiber-optic isolators designed for $1550$-$\nano\meter$ operation are vulnerable to picosecond pulses at $1061~\nano\meter$~\cite{ruzhitskaya2021, ponosova2025}. Exposure to a low average power of just $17~\milli\watt$ caused a temporary isolation reduction of $19.5~\deci\bel$, while more powerful sub-nanosecond pulses led to irreversible component degradation~\cite{ponosova2025}. These results show that a countermeasure effective against $1550$-$\nano\meter$ cw laser attacks, such as an additional protective isolator, may be insufficient against more sophisticated LDA scenarios and may compromise the components of the QKD located behind it~\cite{ponosova2022}. The relevance of the $1061~\nano\meter$ wavelength for light-injection attacks is underscored by independent characterizations, identifying it as a high-risk band due to reduced attenuation in common source configurations~\cite{tan2025}.
	
%In this work, we extend our investigation by examining the effects of PL radiation at $1061~\nano\meter$ on fiber-optic attenuators. We study the impact of different laser generation modes on several types of attenuators widely used in commercial QKD systems. By comparing our results with previously reported data for cw lasers~\cite{huang2020}, we demonstrate a vulnerability: the use of PL radiation significantly lowers the optical damage threshold of the components compared to direct cw attacks. This effect allows an adversary to compromise the device using substantially lower average power than previously thought necessary. The goal of this study is to identify new vulnerabilities to help the development of more robust countermeasures, enhancing the practical security of quantum communication systems.

In this work, we extend our investigation by examining the effects of PL radiation at $1061~\nano\meter$ on fiber-optic attenuators designed for operation at $1550$-$\nano\meter$. This study is a step towards the development of more robust countermeasures, defending all the possible LDA scenarios. We study the impact of differentlaser pulsing regimes on several types of attenuators used in commercial QKD systems. Our study identifies the types of attenuators that are resilient against $1061~\nano\meter$ PL LDA and demonstrates a new potential vulnerability for the other types. 
Specifically, it shows that PL pre-exposed fixed attenuators with an absorption element may exhibit reduced attenuation at $1550~\nano\meter$ within their operational powers, enabling an adversary to compromise the device before its installation in the QKD transmitter.
	
The article is organised as follows. In \Cref{sec:exp_meth}, we explain the experimental setup and methodology employed for testing the fiber-optic attenuators. The measurement results are detailed in \cref{sec:results}. In \cref{sec:countermeasure}, we delve into the effects caused by this attack and explore potential countermeasures. We conclude in \cref{sec:conclusion}.
	
	\section{Experimental methodology}
	\label{sec:exp_meth}

	\subsection{Experimental setup}
	\label{sec:setup}

\begin{figure}
			\includegraphics{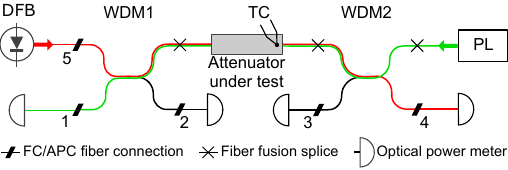}
			\caption{Experimental setup. DFB, $1550$-$\nano\meter$ distributed-feedback laser; WDM, wavelength-division multiplexer; PL, pulsed laser; TC, thermocouple fixed on the attenuator housing. Light directions with minimal loss in setup are denoted by green (light gray) for $1061~\nano\meter$ and red (dark gray) for $1550~\nano\meter$.}
			\label{fig:setup1}
		\end{figure}

		The experimental setup shown in \cref{fig:setup1} simulates a potential hacking scenario where an eavesdropper disrupts the attenuator by injecting PL radiation. The direction of the PL radiation is opposite to that of the attenuator's transmission path. To track the attenuation coefficient of the component under test at the QKD operation wavelength, a $1550$-$\nano\meter$ distributed feedback laser diode (DFB; Gooch \& Housego AA1406) imitates the QKD source signal. It emits continuously, delivering a power of $45.8~\milli\watt$.
		
		Two dense wavelength-division multiplexers (WDMs) are placed on either side of the attenuator under test, enabling simultaneous transmission of radiation from DFB and PL through the sample. The red (dark gray) and green (light gray) lines in~\cref{fig:setup1} denote the $1550~\nano\meter$ and $1061~\nano\meter$ wavelengths, respectively. The $1550~\nano\meter$ light transmitted through the sample is measured at port~4 using an optical power meter (Thorlabs PM400 with an S154C sensor). This measurement is compared to the input power, corrected for loss introduced by the setup components, allowing us to estimate the attenuation and track its changes during the experiment. Monitoring port~2 with a second optical power meter (Thorlabs PM20CH) enables continuous verification of the operational stability of the DFB laser and the WDMs throughout the measurements.
		
		\begin{figure}[b]
			\includegraphics{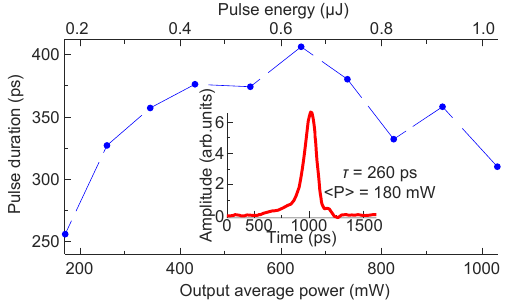}
			\caption{The dependence of average output power and pulse duration of PL in single-pulse regime. Inset shows the pulse shape.}
			\label{fig:calibrationPL}
		\end{figure}

		The attack implementation employs the PL setup we previously used~\cite{ponosova2025}. This setup was developed at the Laboratory of Active Media for Solid-State Lasers at Prokhorov General Physics Institute of Russian Academy of Sciences \cite{trikshev2016, Zhluktova2020}. The system consists of a master oscillator and two ytterbium-doped fiber amplifier (YDFA) stages. The laser generates $1061$-$\nano\meter$ single pulses or pulse trains at a $1~\mega\hertz$ repetition rate, chosen to balance two key factors: significant single-pulse energy for initiating damage and failing to reach material ablation~\cite{wood2003}, with adjustable average output power from $170$ to $1030~\milli\watt$. The pulse duration exhibits a nonlinear dependence on output power, shown in \cref{fig:calibrationPL}, varying between $260$ and $410~\pico\second$ due to nonlinear effects in the active amplifier media (a detailed overview of the PL system is provided in~\cite{ponosova2025}). Its average output power is monitored at port~3 using an optical power meter (Ophir Orion TH with an Ophir~3A sensor). We monitor the power transmitted through the tested attenuator at port~1 of the experimental setup with an optical power meter (Thorlabs PM20CH).

Additionally, we monitor the temperature on the surface of the sample during the test. For this purpose, a Cr/Al thermocouple with temperature meter (Center 301) is attached to the test sample's housing using thermally conductive silicone organic paste and aluminum tape. 
	
		\subsection{Test procedure}
		\label{sec:procedure}
		
We conduct a comprehensive characterization of the experimental setup before attenuator tests. This includes assessing the loss introduced by each element of the experimental setup to varying PL power levels.
% This includes evaluating the response of all setup components to varying PL power levels and assessing the loss introduced by each element of the experimental setup. 
This calibration is essential for ensuring accurate measurements, particularly owing to the nonlinear changes in WDM splitting ratio with increasing PL power.
		
\begin{figure}
	\includegraphics{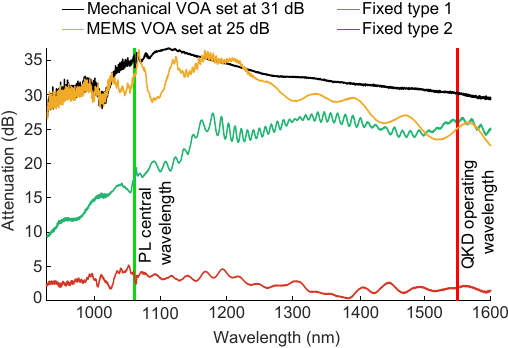}
	\caption{ Initial transmission spectra of the testing attenuators. The spectra show the intrinsic attenuation profile of each sample. The red and green vertical lines denote the operating wavelength of the QKD system and the central wavelength of the PL used in tests, respectively.}
	\label{fig:spectrum}
\end{figure}

The experimental procedure begins with preliminary sample characterization via transmission-spectrum measurements using a custom supercontinuum source (NGO Soliton) and an optical spectrum analyzer (Yokogawa AQ6370D). The results, shown in~\cref{fig:spectrum}, allow us to compare sample transparency at the attack wavelength ($1061~\nano\meter$, green line) and the typical QKD operating wavelength ($1550~\nano\meter$, red line).
Note that this characterization serves for preliminary comparison of the tested samples. Their comprehensive security analysis would require measurements across a broader spectral range, as demonstrated in~\cite{tan2025}.
		
We then calibrate each attenuator and install it in the experimental setup. For variable optical attenuators (VOAs), we set attenuation levels representative of typical QKD system operation. Each sample undergoes sequential testing in two PL regimes. Initially, we employ the single-pulse mode.
%Each attenuator is then calibrated and installed in the experimental setup. For variable attenuators, we set attenuation levels representative of typical QKD system operation. Each sample was tested sequentially in two PL regimes. First, the single-pulse mode is employed. %using pulses with a duration of $260$ and $410~\pico\second$, a repetition rate of $1~\mega\hertz$, and an average output power ranging from $132$ to $1030~\milli\watt$. \APc{delete "using pulses with a duration of $260$ and $410~\pico\second$, a repetition rate of $1~\mega\hertz$, and an average output power ranging from $132$ to $1030~\milli\watt$"} 
If no significant changes are observed under this condition, testing proceeds to the multi-pulse mode. In this regime, the $1~\mega\hertz$ repetition rate corresponds to the frequency between trains of pulses, with each containing up to seven pulses. %, with average output power from $132$ to $1030~\milli\watt$ \APc{delete , with average output power from $132$ to $1030~\milli\watt$}. 
The attenuation of tested samples at $1550~\nano\meter$ is continuously monitored throughout the exposure to PL.
		
The testing in each laser mode starts with sample irradiation using the attack PL at an initial average power of $170~\milli\watt$ for at least five minutes. If no significant change in attenuation is detected, the average power is increased stepwise according to~\cref{fig:calibrationPL}. The sample is exposed for $5~\minute$ at each level until attenuation changes are observed or the maximum available power of $1030~\milli\watt$ is reached. Subsequently, the PL is turned off, allowing a $10$--$20~\minute$ stabilization period for residual effect verification. After stabilization, the sample's attenuation is re-measured to quantify any permanent changes. This measurement is first carried out using a low-power DFB laser at $1550~\nano\meter$ to establish the baseline attenuation, and then repeated under an increased cw power of $500~\milli\watt$ from a lab-made amplified spontaneous emission (ASE) source to verify the sample’s stability. If no changes are observed after the single-pulse tests, the procedure is repeated in the multi-pulse regime. Fixed attenuators are additionally characterized using a $2$-$\watt$ high-power cw laser operating at $1550~\nano\meter$, described in~\cite{huang2020,ponosova2022}.
		
The attack is considered successful when the component's attenuation at its operating wavelength of $1550~\nano\meter$ decreases by at least $1~\deci\bel$ under PL exposure. This attenuation reduction corresponds to a $26\%$ increase in the quantum channel's mean photon number, potentially enabling key interception~\cite{sajeed2015, huang2019, zheng2019}. Conversely, attenuation increases exceeding $3~\deci\bel$ are classified as denial-of-service, as they disrupt key generation~\cite{huang2020}.
	
\section{Experimental results}
\label{sec:results}

The study evaluates four types of optical attenuators that are currently in use in practical QKD systems. VOAs with blocking elements and fixed attenuator with gap loss demonstrate stability to PL impact across all regimes. Micro-electro-mechanical systems (MEMS)-based attenuators exhibit vulnerability, with attenuation reduction up to $3.8~\deci\bel$ when subjected to PL in a multi-pulse regime. %Fixed attenuator with gap loss exhibits stability to PL attack. 
Fixed attenuators with an absorption element, following preliminary PL exposure, exhibited a temporary attenuation reduction of up to $7~\deci\bel$ when illuminated with a $1~\watt$ cw laser at an operating wavelength of $1550~\nano\meter$. A summary of the detailed test results is provided in \cref{tab:results} and discussed below. Notably, in contrast to the cw laser attack scenario reported in~\cite{huang2020}, no sample heating beyond the specified operational temperature limits was observed.

\begin{table*}
	\vspace{-0.7em} % compensates for REVTeX layout bug
	\caption{Optical damage results for all the tested attenuators.}
%		\resizebox{\linewidth}{!}{ % uncomment to shrink table
	\begin{tabular}[t]{lcccccccc}
	\hline\hline
		     \makecell{Attenuator\\ type} & Model & \makecell{Used in\\ QKD\\ system(s)} & \makecell{Number of  \\ damaged \\ /tested \\ samples } & 
		     \makecell{Damage\\ in single-\\pulse\\ regime} & 
		     \makecell{Damage\\ in multi-\\pulse\\ regime} & 
		     \makecell{Highest\\ attenuation\\ change at\\ $1550~\nano\meter$ ($\deci\bel$)} &
		     \makecell{Operating\\ temperature\\ range ($\celsius$) \\} &
		     \makecell{Maximum\\ temperature\\ in tests ($\celsius$) \\} \\ \hline
			\makecell[l]{VOA with\\ blocking\\ element} & \makecell{OZ Optics BB-\\700-11-1550-8/125-\\P-60-3A3A-1-1-LL} & \cite{jouguet2013a} & $0/2$    &  $-$   &  $-$  &  0  & $-30$~to~$+70$ & $+58$   \\ 
		        MEMS VOA                      & undisclosed\footnote{\label{ft:withheld}Information withheld at the request of QKD manufacturer.} & undisclosed\footref{ft:withheld} &  $1$\footnote{\label{ft:defect}The effect may be caused by manufacturing variations.}$/3$    &  $-$   &  $+$  &   $-3.8$, permanent   & $-5$~to~$+75$  & $+73$   \\ 
		        Fixed type~1                  & undisclosed\footref{ft:withheld} & undisclosed\footref{ft:withheld}  &  $3/3$    &  $+$   &  $+$  &   $-7$\footnote{\label{ft:measure}Measured at $1~\watt$ with cw high-power laser operating at $1550~\nano\meter$.}, temporary   & $-40$~to~$+85$ & $+45$ \\ 
		        Fixed type~2                  & undisclosed\footref{ft:withheld} & undisclosed\footref{ft:withheld} &  $1/1$    &  $-$   &  $+$  &   $-0.8$\footref{ft:measure}, temporary   & $-40$~to~$+85$ & $+40$ \\
	\hline\hline
	\end{tabular}
%	} % uncomment to shrink table
	\label{tab:results}
\end{table*}

The study evaluates four types of optical attenuators that are used in practical QKD systems (see \cref{tab:results}). Variable optical attenuators (VOAs) with blocking elements and fixed attenuator with gap loss demonstrate stability to PL impact across all regimes. Micro-electro-mechanical systems (MEMS)-based attenuators exhibit vulnerability, with attenuation reduction up to $3.8~\deci\bel$ when subjected to PL in a multi-pulse regime. Fixed attenuators with an absorption element, following preliminary PL exposure, exhibited a temporary attenuation reduction of up to $7~\deci\bel$ when illuminated with an $1$-$\watt$ cw laser at an operating wavelength of $1550~\nano\meter$. Notably, in contrast to the cw laser attack scenario reported in~\cite{huang2020}, no sample heating beyond the specified operational temperature limits was observed.

\subsection{Variable attenuator with blocking element}
\label{sec:mech}
		
\begin{figure}[b]
	\centering
	\includegraphics[width=\linewidth]{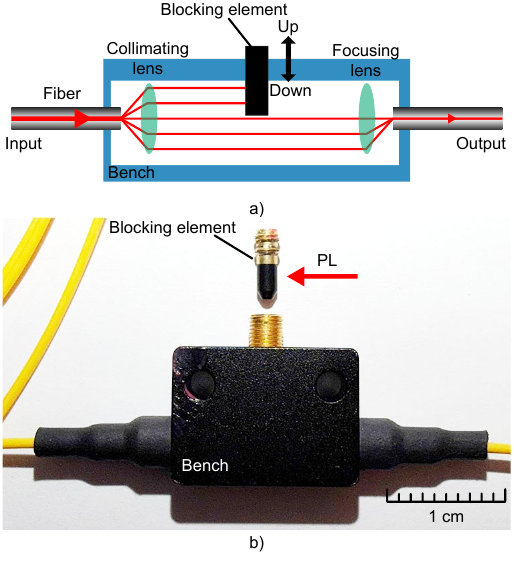}
	\caption{Manual VOA with blocking element. (a)~Simplified cross-section (not to scale). Red arrows denote light propagation direction. (b)~Photograph of tested sample.}
	\label{fig:manualVOA}
\end{figure}

A mechanical VOA operates by partially blocking a collimated optical beam with an opaque element, as shown in~\cref{fig:manualVOA}. The device is evaluated at a fixed attenuation setting of $31~\deci\bel$. Exposure to sub-nanosecond pulsed radiation, in both single-pulse and multi-pulse regimes and across the full available range of average power, resulted in no observable changes in its performance. Additional tests conducted at a reduced attenuation of $20~\deci\bel$, with exposure times increased up to $30~\minute$, further confirmed the component's resistance to the attack.

The thermal effect from PL proved significantly weaker than from cw exposure. During testing, the maximum housing temperature reached $+58\celsius$, remaining within the component's operational range specified in \cref{tab:results}. For comparison, reference~\cite{huang2020} reported that under cw laser attack, another sample of the same attenuator heated to $+234\celsius$, exceeding its rated specifications and causing visible housing damage.
		
Thus, the mechanical attenuator demonstrated complete resistance to sub-nanosecond laser irradiation within the tested parameters.

\subsection{MEMS-based variable attenuator}
\label{sec:MEMS}
		
\begin{figure}
	\includegraphics{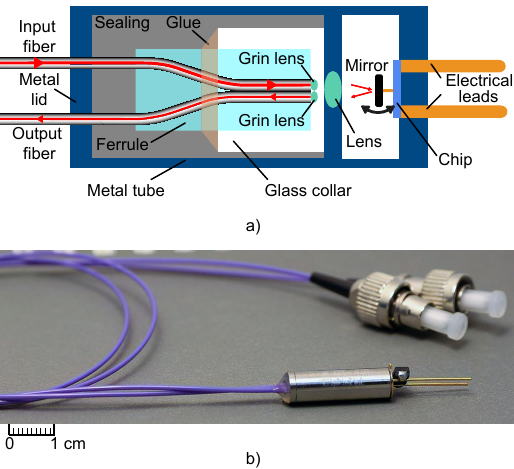}
	\caption{MEMS variable optical attenuator. (a)~Simplified cross-section showing main elements (not to scale). Red arrows denote light propagation direction. (b)~Photograph of the test sample.}
	\label{fig:MEMS}
\end{figure}

We test three samples of the MEMS variable attenuator of the same model as in~\cite{huang2020}. The device operates by adjusting the angle of an internal dielectric mirror, thereby controlling the coupling efficiency between the input and output fibers over an attenuation range of approximately $1$ to $31~\deci\bel$ (\cref{fig:MEMS}). The attenuation is controlled by applying dc voltage from power supply (GW~Instek SPS-1820). In our tests we set attenuation of every sample to approximately $25~\deci\bel$.
		
\begin{figure}
	\includegraphics{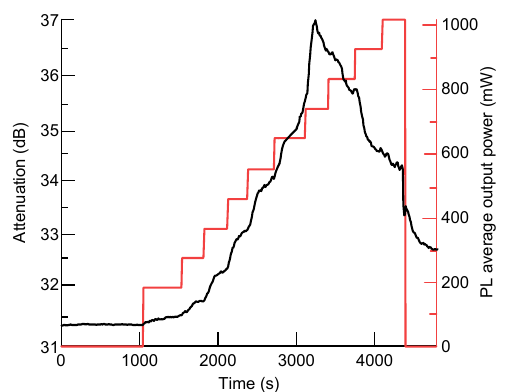}
	\caption{Variation of the MEMS attenuation coefficient (black line) depending on the incident power of PL (red or gray line) in single-pulse regime.}
	\label{fig:MEMSsinglepulse}
\end{figure}

Under single-pulse exposure, we observe an increase in attenuation with increasing attack power, as shown in \cref{fig:MEMSsinglepulse}. This effect is slowly reversible: within a few days after the exposure, the attenuator characteristics recover to the initial values. We attribute this to a relatively high single-pulse energy (up to $1~\micro\joule$ at the average power of $1030~\milli\watt$) likely inducing temporary nonlinear effects or short-lived defect formation (color centers) in the lens or dielectric mirror materials. This increases light scattering and consequently causes temporary attenuation growth~\cite{Manenkov2014}.
		
\begin{figure}
	\includegraphics{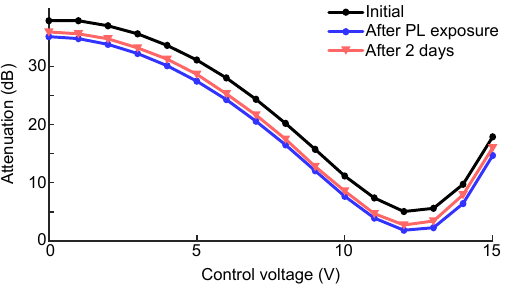}
	\caption{Alteration in the attenuation of MEMS VOA following testing in the multi-pulse exposure regime.}
	\label{fig:MEMSmultipulse}
\end{figure}

In a subsequent experiment, multi-pulse exposure induced an irreversible attenuation decrease of up to $3.8~\deci\bel$ across the operational voltage range. This effect was observed in only one of the three samples tested. Two other samples behaved similarly to the single-pulse mode exposure and recovered after the test. As shown in \cref{fig:MEMSmultipulse}, the change persisted for $48~\hour$ post-exposure, confirming permanent device modification.
Although all samples were brand new, the damaged sample initially deviated from specifications, exhibiting a minimum attenuation of $5~\deci\bel$  rather than the expected $1~\deci\bel$. This suggests a pre-existing manufacturing defect, which likely acted as a localized absorption site. Under multi-pulse irradiation, such a defect might reduce the laser-induced damage threshold and initiate the observed attenuation reduction. The maximum recorded temperature during the experiment approached the operational limit of $+73\celsius$ (\cref{tab:results}).
		
\begin{figure}
	\includegraphics{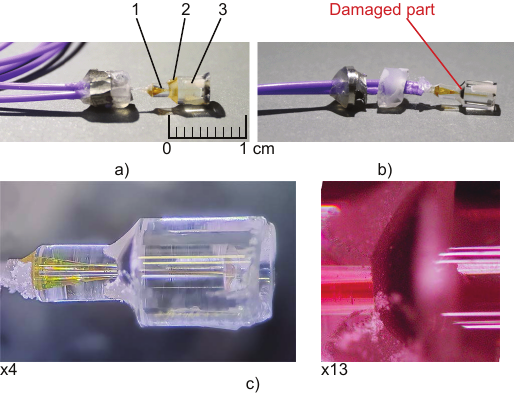}
	\caption{Internal MEMS components after attack by (a)~cw laser (damaged in~\cite{huang2020}) and (b)~PL. 1, ferrule; 2, glue; 3, glass collar. Darkening area indicates damage after exposure to PL. (c)~Optical microscope images of the damaged area after exposure to PL.}
	\label{fig:photodamage}
\end{figure}

A detailed microscopic examination of the PL-exposed sample revealed a damage pattern different from that of the cw-laser-damaged attenuator we previously reported~\cite{huang2020}. A pronounced darkening area is visible at the adhesive-filled interface between the ferrule and the surrounding glass collar (\cref{fig:photodamage}). The damaged area is a dark spot originating near the central axis and distributing radially. This observation suggests that the optic glue acted as the primary absorption medium and failure point under high-power pulsed exposure. This degradation led to the disturbance of the optical axis of the device and, as a result, changed the attenuation uniformly through the full voltage range (\cref{fig:MEMSmultipulse}). No other signs of thermal or mechanical damage were found elsewhere within the component. The possible damage mechanism is the thermal decomposition and carbonization of the adhesive, initiated by the intense power of the sub-picosecond pulses. This may be induced by absorption by impurities within the adhesive, leading to cumulative heating during the multi-pulse exposure and leading to disturbance of the optical part of the device~\cite{smith2008, amina2019}. 

\subsection{Fixed attenuators}
\label{sec:fixed}
	
\begin{figure}[b]
	\includegraphics{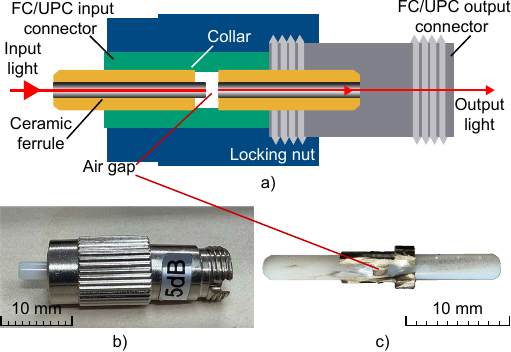}
	\caption{Fixed attenuator with gap-loss. (a)~Simplified cross-section (not to scale). Red arrows denote light propagation direction. (b)~Photograph of the attenuator. (c)~The internal elements under the section of metal housing.}
	\label{fig:fixedgap}
\end{figure}

\begin{figure}
	\includegraphics{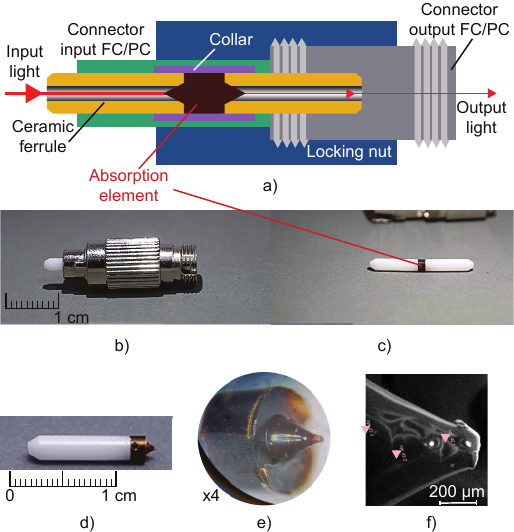}
	\caption{Fixed attenuator with absorption element. (a)~Simplified cross-section (not to scale). Red arrows denote light propagation direction. (b)~Photograph of the attenuator. (c)~The internal component with the metal housing removed. (d)~Magnified view of one of the ferrules with the absorbing element. (e)~Optical microscope (Altami SM~1065) image of the absorbing element's tip. (f)~Scanning electron microscope (Quattro~S) image of the same tip. Red triangles mark the areas where elemental analysis was performed.}
	\label{fig:fixed}
\end{figure}

We test fixed attenuators of two types: absorption-base (samples 1--3) the same as in~\cite{huang2020} and gap-loss (sample~4), which reduces optical power via an air gap between fibers~\cite{azadeh2009} (\cref{fig:fixedgap}). The latter exhibits stability under PL exposure in both regimes, with variation in attenuation less than the success attack threshold of $1~\deci\bel$. The absorption-based design uses a special element positioned between input and output fibers, as schematically illustrated in \cref{fig:fixed}(a). Under exposure to PL operating in the single-pulse generation mode, their  attenuation value fluctuates by just $\pm 1~\deci\bel$. At the maximum average output power of $1030~\milli\watt$, the samples heat up to $45\celsius$. However, no irreversible damage or performance degradation occurs. Similarly, subsequent samples' exposure to PL in the multi-pulse mode does not cause significant changes in the attenuation coefficient.
		
%The initial stage of the investigation is conducted in the single-pulse generation mode. Change in the attenuation value that varied within $\pm 1~\deci\bel$ has been recorded. At the maximum average output power of $1030~\milli\watt$, heating of the samples to $43\celsius$ is observed; however, no signs of irreversible damage or performance degradation are detected. Subsequently, the samples are tested in the multi-pulse mode, but no significant changes in the attenuation coefficient have been recorded.

However, we have found that preliminary PL irradiation induces latent changes in the attenuator material. These changes manifested under subsequent exposure to cw at the wavelength of $1550~\nano\meter$. Initial testing with a $0.5~\watt$ ASE source at $1550~\nano\meter$ revealed this effect: in sample~1, the attenuation temporarily dropped from $25.1$ to $19.3~\deci\bel$. To quantify the vulnerability further, we employed a high-power cw laser~\cite{huang2020, ponosova2022}. We focused on the $1~\watt$ power level, as the manufacturer guarantees attenuator characteristics up to and including this power limit. The results show that after the preliminary PL treatment, a cw power of $1~\watt$ is sufficient to induce a significant attenuation reduction of $1.9$--$7~\deci\bel$ and increases with applied laser power~(\cref{tab:fixed}). This reduction exceeds the effect achieved by direct high-power attacks reported in prior work~\cite{huang2020}.

The direct LDA required a $4$-$\watt$ cw laser at $1550~\nano\meter$ to achieve a $1$--$2~\deci\bel$ attenuation reduction. Our method achieves a comparable or greater impact at a substantially lower cw power ($1~\watt$) by first ``priming" the component with PL. This approach is not only more efficient but also more covert, as the preparatory PL stage can be executed at power levels that do not trigger immediate failure or obvious disturbance in system work.		
		
\begin{table*}
	\vspace{-0.7em} % compensates for REVTeX layout bug
	\caption{Changes in absorption-base fixed attenuators at $1550~\nano\meter$ after PL exposure.}
	\begin{tabular}[t]{lccccccc}
		\hline\hline
		\multirow{2}{*}{Type} & 
		\multirow{2}{*}{\makecell{Sample\\ number}} & 
		\multirow{2}{*}{\makecell{Specification\\ attenuation}} & 
		\multirow{2}{*}{\makecell{Initial\\ attenuation ($\deci\bel$)}} & 
		\multirow{2}{*}{\makecell{Attenuation after PL\\ exposure ($\deci\bel$)}} & 
		\multicolumn{3}{c}{\makecell{Attenuation under \\ cw exposure at $1550~\nano\meter$ ($\deci\bel$)}} \\  
		\cline{6-8}
		& & & & & 
		\makecell{at $0.5~\watt$} & 
		\makecell{at $1~\watt$} & 
		\makecell{at $2~\watt$} \\ 
		\hline 
		\multirow{3}{*}{1 (absorption element) \rotatebox[origin=c]{0}{$\Bigg\{$}} & 
		$1$ & $25$ & $25.6$ & $24.1$ & $19.3$ & $18.6$ & $17.8$ \\
		&$2$ & $25$ & $25.6$ & $22.1$ & $22.6$ & $23.7$ & $17.4$ \\
		&$3$ & $15$ & $15.4$ & $15.3$ & $14.4$ & $13.0$ & $13.1$ \\
		2 (gap loss) & $4$ & $5$ & $4.9$ & $5.5$ & $4.4$ & $4.1$ & $4.7$ \\          
		\hline\hline
	\end{tabular}
	\label{tab:fixed}
\end{table*}
		
To understand the reasons for this behavior, we perform a detailed analysis of the attenuator's internal structure, paying particular attention to the absorbing element, as illustrated in \cref{fig:fixed}(b)--(f). Its elemental composition is determined by an energy-dispersive X-ray spectroscopy using a spectrometer (Bruker Quantax). The absorbing element has a cylindrical shape with conical ends [see \cref{fig:fixed}(c)--(f)] and is a composite material comprising carbon, oxygen, iron, and nickel. This finding is of interest when compared to the data presented in~\cite{huang2020}, where exposure to cw laser radiation led to thermal destruction of the absorbing element. In our case, however, despite the intense pulsed exposure, no visible damage to the samples is observed.

Based on the obtained data, we propose the following two-stage damage mechanism.
\begin{itemize}
	\item[--] At the first stage, sub-nanosecond pulses from the PL at $1061~\nano\meter$, through nonlinear processes, cause chemical bond disruption within the carbon-based matrix with metallic inclusions (iron, nickel)~\cite{Manenkov2014, kallepalli2005, bityurin2005}. At a high repetition rate of $1~\mega\hertz$, this leads to an accumulation of latent defects, such as microcracks or oxidized carbon regions—which serve as new absorption centers~\cite{urech2010}.
	\item[--]  At the second stage, the subsequent cw radiation at $1550~\nano\meter$ is intensely absorbed by these newly formed defects, further degrading the attenuating properties of the material~\cite{qingliang2023}. This process accelerates the attenuation drop (\cref{tab:fixed}).
\end{itemize}

\section{Discussion and countermeasures}
\label{sec:countermeasure}

Our component-level analysis reveals varied responses of fiber-optic attenuators to PL exposure. This component-level analysis provides preliminary insights into the impact of PL on different attenuator types, serving as a first step for comprehensive QKD security assessments.
		
The investigation was motivated by the risk of cascading vulnerabilities. Our previous work~\cite{ruzhitskaya2021, ponosova2025} showed that optical isolators, the primary defense for QKD sources, can be compromised by $1061$-$\nano\meter$ PL attacks, allowing radiation to reach subsequent components such as attenuators. We show that mechanical VOAs with a blocking element and fixed gap-loss attenuators maintain integrity under PL exposure. MEMS attenuators exhibit complex behavior, leading to temporary increases or permanent decreases in attenuation. The most significant vulnerability was found in fixed attenuators with absorbing elements, where PL changes the absorption material and leads to a decrease in attenuation at operating wavelength. 

This potential for a cascading vulnerability, where an attack first compromises the optical isolator or optical power limiter~\cite{peng2024} and then the attenuator, could create a two-pronged security threat. First, PL attack may compromise the system's defense against light-injection attacks by significantly degrading the isolation performance of the optical isolator. As demonstrated in our previous study~\cite{ponosova2025}, this attack can decrease the isolation coefficient of the isolator by $19.5~\deci\bel$ and disturb the recommended security threshold of $60$--$80~\deci\bel$ necessary to oppose Trojan-horse attacks~\cite{lucamarini2015}. Second, the same PL may reach the fixed attenuator behind the isolator and induce hidden damage. It leads to an open side channel for reduction of attenuation by $1.9$--$7~\deci\bel$ at a power level of $1~\watt$ at $1550~\nano\meter$. This reduction could lead to an exponential increase in the mean photon number, potentially facilitating light-injection attacks~\cite{sajeed2015, huang2019, zheng2019, scarani2004}. 

Potential countermeasures include inserting a narrow-band wavelength-division multiplexer to filter the quantum channel~\cite{han2023}, using a circulator with a fiber Bragg grating for combined isolation and narrow filtering~\cite{tan2025, comandar2020}, or employing microstructured photonic bandgap fibers to block a broad range of excess wavelengths~\cite{McGarry2024}.

Further research should focus on testing the full component chain (e.g., attenuator + isolator) to verify cascading effects experimentally. Additionally, collecting statistical data across more samples is necessary to establish reliable thresholds for attenuation degradation and to support the integration of such components into standardized QKD security proofs.

\section{Conclusion}
\label{sec:conclusion}

In this work, we have investigated the effects of PL damage attacks on four types of fiber-optic attenuators. We have found that the variable attenuators with the mechanical blocking element and fixed attenuator with the gap-loss design are resilient to such attacks. The MEMS-based attenuator has experienced a permanent $3.8$-$\deci\bel$ reduction in attenuation, while the fixed attenuator with absorbing element showed a greater reduction of up to $7~\deci\bel$ at its operating wavelength. Unlike previously studied direct damage scenarios, we demonstrate that initial exposure to PL can change properties of internal material, rendering it more vulnerable to a subsequent, lower-power attack using a standard $1550$-$\nano\meter$ cw laser. This finding highlights a sophisticated threat vector where a two-stage attack can bypass standard defenses. Consequently, our results emphasize that current countermeasures are insufficient and call for more comprehensive component certification protocols that account for a wider spectrum of laser damage scenarios, including pulsed and cascading attacks.

\bigskip
\textbf{List of abbreviations:}
QKD, quantum key distribution; LDA, laser-damage attack; cw, continuous-wave; PL, pulsed laser; DFB, distributed-feedback laser; WDM, wavelength-division multiplexer; TC, thermocouple;  YDFA, ytterbium-doped fiber amplifier; ASE, amplified spontaneous emission; VOA, variable optical attenuator; MEMS, micro-electromechanical system.

\section*{Declarations} 

\textbf{Availability of data and materials:}
Raw experimental data and calculations can be obtained from the corresponding author upon a reasonable request.

\medskip
\textbf{Competing interests:}
The authors declare no competing interests.

\medskip
\textbf{Funding:}
The work was supported by the Russian Science Foundation (project №23-79-30017). D.T.\ and V.M.\ acknowledge funding from the Galician Regional Government (consolidation of research units: atlanTTic and own funding through the ``Planes Complementarios de I+D+I con las Comunidades Autonomas'' in Quantum Communication), the Spanish Ministry of Economy and Competitiveness (MINECO); the Fondo Europeo de Desarrollo Regional (FEDER) through grant PID2024-162270OB-I00, MICIN with funding from the European Union NextGenerationEU (PRTR-C17.I1), the ``Hub Nacional de Excelencia en Comunicaciones Cu{\'a}nticas'' funded by the Spanish Ministry for Digital Transformation and the Public Service and the European Union NextGenerationEU, the European Union's Horizon Europe Framework Programme under Marie Sk\l{}odowska-Curie grant 101072637 (project QSI) and project ``Quantum Security Networks Partnership'' (QSNP; grant 101114043), and the “Programa de Cooperaci{\'o}n Interreg VI-A Espa\~na–Portugal” (POCTEP) 2021–2027 through the project QUANTUM\_IBER\_IA. A.H.\ and T.X.\ acknowledge funding from the National Natural Science Foundation of China (No. 62371459).
% A.Z.,\ G.T.,\ and V.K.\  acknowledge funding from the Russian Science Foundation 

\medskip
\textbf{Authors' contributions:}
D.R.,\ I.Z.,\ A.P.,\  F.U.,\ and V.K.\ conducted the experiment. I.Z.,\ A.Z.,\ G.T.,\ and V.K.\ assembled the high-power laser. D.R.,\ I.Z.,\ D.T.,\ A.P.,\ T.X.,\ A.H.,\ and V.K.\ analyzed the data. D.R.\ and K.M.\ conducted microscopic studies.  D.R.,\ D.T.,\ I.Z.,\ and A.P.\ wrote the article with help from all authors. V.K.,\ V.T.,\ and V.M.\ supervised the project.

\medskip
\textbf{Acknowledgments:}
We thank our industry partners for providing us device samples. The microscopic examinations were performed at the Advanced Imaging Core Facility (AICF) at Skolkovo Institute of Science and Technology. 

\medskip
\textbf{Ethics approval and consent to participate:} 
Not applicable.

\medskip
\textbf{Consent for publication:}
All authors have approved the publication. The research in this work did not involve any human, animal or other participants.

\def\bibsection{\medskip\begin{center}\rule{0.5\columnwidth}{.8pt}\end{center}\medskip} % Redefines bibliography separator to single-column. This reduces chances of float placement bugs in the last page.
\bibliography{library}

\end{document}